\documentclass[aps,pra,superscriptaddress,nofootinbib]{revtex4-2}
\usepackage{leftindex,tensor,mhchem}
\usepackage{amsmath,amsthm,amsfonts,amssymb,amscd,dsfont} 
\usepackage{indentfirst}
\usepackage{graphicx}
\usepackage{color}
\usepackage[american]{babel}
\usepackage{float}
\usepackage[dvipsnames]{xcolor}
\usepackage[colorlinks=true,citecolor=ForestGreen,linkcolor=magenta,urlcolor=Mulberry,hyperindex]{hyperref}
\usepackage{cleveref}
\usepackage{mathtools}
\usepackage{bm}
\usepackage{bbold}
\usepackage{units}
\usepackage{appendix}
\usepackage{comment}
\usepackage{etoolbox}
\usepackage{soul}

\usepackage{array}

\makeatletter
\def\@email#1#2{%
 \endgroup
 \patchcmd{\titleblock@produce}
  {\frontmatter@RRAPformat}
  {\frontmatter@RRAPformat{\produce@RRAP{*#1\href{mailto:#2}{#2}}}\frontmatter@RRAPformat}
  {}{}
}%
\makeatother

\newcommand{\id}{\mathbb{1}}

\newtheorem{theorem}{Theorem}
\newtheorem*{theorem*}{Theorem}
\newtheorem{corollary}{Corollary}

\newcommand{\rcqi}{Institute of Physics, Slovak Academy of Sciences, Dubravska cesta 9, Bratislava 845 11, Slovakia}
\newcommand{\cft}{Center for Theoretical Physics, Polish Academy of Sciences, Aleja Lotnik\'{o}w 32/46, 02-668 Warsaw, Poland}

\begin{document}

\title{Higher-order incompatibility improves distinguishability of causal quantum networks}

\author{Nidhin Sudarsanan Ragini}
\affiliation{\rcqi}

\author{Sk Sazim$^{*,}$}
\email{Corresponding author: sk.sazimsq49@gmail.com}
\affiliation{\cft}

\date{\today}

\begin{abstract}
Higher-order quantum theory deals with causal quantum processes, described by quantum combs, and test procedures, described by quantum testers, ``measuring" these processes. In this work, we show that ``jointly non-implementable" or incompatible quantum testers perform better in distinguishability tasks than their compatible counterparts. To demonstrate our finding, we consider a specific two-party game based on distinguishing quantum combs. We show that the player does better at winning the game when they have exclusive access to incompatible testers over compatible ones.  Moreover, we show that, using the resource theoretic measure convex weight,  any general quantum resource present in testers is resourceful in quantum comb exclusion tasks. These investigations generalise, respectively, an earlier finding that incompatibility of quantum observables to be a bona fide resource in quantum state distinguishability tasks and another finding that any resource present in observables result in improved performance in state exclusion or antidistinguishability tasks.
\end{abstract}

\maketitle


\section{Introduction}
\noindent Historically, one of the perplexing avenues of quantum {theoretical} studies has been the notion of non-commuting observables and their uncertainty relations, manifesting {as ``{non-joint-measurability}"} as well as error-disturbance trade-offs \cite{Heinosaari_Ziman_2011}. In the current age of active development and understanding of quantum information processing and quantum computation \cite{Nielsen_Chuang_2010}, this avenue has transformed into studies on joint measurability beyond non-commutativity (See \cite{guhne2023colloquium} for a recent colloquium and \cite{heinosaari2016invitation} for an overview of related aspects) and, subsequently extending this notion to ``{joint implementation}" of other quantum devices beyond quantum measurements \cite{heinosaari2017incompatiblechannels, leppajarvi2024incompatibility, sedlak2016incompatible, heinosaari2016invitation, FrancescoIncompatibility}. This inadmissibility of joint implementation of certain collections of quantum devices is now referred to as ``{quantum incompatibility}" {pertaining to}  these collections \cite{guhne2023colloquium,heinosaari2016invitation}. 
Moreover, the incompatibility of quantum observables can be understood as a bona fide resource that is consumed in certain quantum protocols, for example, in enhancing secure key rates in quantum cryptographic protocols \cite{RevModPhys.74.145}.

Recently, along the above lines, it has been found that for a particular quantum state discrimination game,  all incompatible collections of observables,  described as positive operator-valued measures (POVMs), result in an advantage over compatible collections \cite{Skrzypczyk2019, teikoIncWit}. This result is similar to those that found all entangled states improve performance of channel distinguishability tasks \cite{allEntanglUsefChannelDiscr,allEntanglUsefChannelDiscr2}.
The aforementioned result on incompatible observables has an important operational significance: A given set of measurement observables are incompatible if and {only if} they provide advantage in such discrimination games over compatible ones \cite{Skrzypczyk2019,teikoIncWit}. Subsequently, these results garnered much attention and sparked studies within the  resource theoretic perspective of quantum objects \cite{PhysRevX.9.031053, Designolle_2019, Oszmaniec_2019,RobustnessUola+ConicProgramm, PhysRevA.109.042403,PhysRevLett.132.150201, FrancescoProgrammbility,PhysRevA.109.012415}. 

In the works which are discussed in this passage, a lot of focus was put on finding the physical implications of resource contained within the quantum objects, specifically quantum states, its transformations, quantum channels and the measurement observables. It has been shown that not only the resources within measurement observables are advantageous in state discrimination tasks, the resources within quantum states play auxiliary roles in discrimination tasks involving of their transformations, specifically, quantum channels and measurement observables \cite{PhysRevX.9.031053}. Moreover, we can observe that three distinctive research directions emerge based on these studies: 1) Identifying and characterising the exact quantum features of measurement observables which are advantageous in quantum state discrimination tasks \cite{Skrzypczyk2019, RobustnessUola+ConicProgramm}, and 2) Studies involving usage of resources present in quantum states for improving  discrimination of quantum observables \cite{PhysRevX.9.031053, PhysRevLett.122.140402}. 
Furthermore, 3) Studies identifying the resourcefulness present in general quantum processes, including channels and higher-order transformations, those which result in advantages in quantum state discrimination tasks \cite{PhysRevA.101.052306}. With regard to the third direction, in \cite{PhysRevA.101.052306} the authors consider higher-order transformations \cite{chiribella2009theoretical} as part of the strategies in their state discrimination games with fixed measurement observables and study the advantages. 

A common feature among all of these studies is that all of them consider resourcefulness of a single quantum object class, among the different classes {which} construct the quantum information processing task. However, we can notice that for discrimination tasks of quantum channels, resources present in both {the} ``{probe state}" and ``{probe observable}" can, in synergy, result in {performance improvements or} advantages. In fact, in Ref.\cite{PhysRevResearch.2.033374}, the authors have shown that the advantages in such tasks  {take form} as the product of the individual advantages arising from each of the test objects.  In addition, such channel discrimination tasks can also be strategised to incorporate intermediate, auxiliary ``{probe transformations}" within the test network (these are referred to as \textit{adaptive strategies} in the literature). When test networks  which implement such strategies are used, the resources present in {the associated component} transformations can also result in advantages, and should be accounted for as well. Unlike in \cite{PhysRevResearch.2.033374}, when we are interested in the collective advantage of a particular test network furnish, rather than the individual contributions of the component probe objects, we can resort to describe the network as a single quantum object. Such descriptions are given by quantum testers \cite{chiribella2008memory,gutoski_quantumstrategies,ziman2008process},  which are unified descriptions of the associated probe objects of a test network. In fact, the framework of quantum testers have been developed to describe not only test networks for discriminating quantum channels with single time-step, but also for general multiple time-step quantum causal processes  which appear themselves as incomplete circuits or networks, also referred to {as} \textit{higher-order quantum processes} or \textit{quantum combs} \cite{chiribella2008quantum,gutoski_quantumstrategies, chiribella2009theoretical}; These combs describe the most general transformations of channels, super-channels and so on. As such, within the premise of quantum discrimination tasks, the central objects to be distinguished  can be ,{in full generality ,} considered as quantum combs and quantum testers describing the associated discrimination networks. This outlook is all-encompassing since quantum states and observables can be considered as special cases of quantum channels with single time steps, which are themselves a special case of quantum combs. 

As partially hinted earlier, the notion of quantum incompatibility is not exclusive to observables. In our work, we consider the incompatibility of quantum testers, which we refer to as ``{higher-order incompatibility}" \cite{sedlak2016incompatible}, and study the resourcefulness of this incompatibility  in the quantum comb discrimination game that we present. Subsequently, we find that incompatible testers are advantageous over compatible collections of testers. Alternatively, our results can be understood as providing the following operational meaning to tester incompatibility. A set of quantum testers are incompatible if and only if they provide advantage over compatible ones in our quantum comb discrimination task. Moreover, this advantage is  quantified in terms of the \textit{robustness of incompatibility} of testers. Thus, in contrast to the works discussed earlier, we can infer that possible non-trivial individual contributions, from the resources in probe states, transformations and observables, to the quantum advantage is {collectively} reflected in the robustness. Moreover, we also study the advantageous roles of tester resources in comb exclusion games using the resource-theoretic measure \textit{convex weight}, which also implies that tester incompatibility being a resource here as well. 
\par 
In Section \ref{II1}, the framework of higher-order quantum theory is sketched, where mathematical characterisations of quantum combs and testers are furnished. Within this section, the notion of incompatibility of testers and an associated characterisation of it is also presented. In Section \ref{III1}, we present the particular discrimination game and our results regarding the advantage furnished by higher-order incompatibility. In Section \ref{IV1}, we characterise the advantage of the incompatibility of testers (or, more generally any quantum resource in it) in comb exclusion games. Finally, we conclude with a discussion in Section \ref{V1}.

\section{Higher-order quantum theory: Incomplete quantum networks}\label{II1}
\noindent {As we shall encounter  a myriad of mathematical objects and associated notations, it is best to clearly associate specific font type or symbol with a specific object. Let us tabulate these notations specifying these associations in Table \ref{tab:my_label}:
\begin{table}[H]
    \centering
    \begin{tabular}{|c|c|}
        \hline
        \textbf{Symbols} & \textbf{Objects}\\
        \hline
        $\mathcal{C}$, $\mathcal{N}$ & quantum channels \\
        \hline
        $\mathcal{C}^{(n)}$ &  quantum combs with $n$-slot \\
        \hline
        $\mathsf{C}^{(n)}$, $\Xi^{(n)}$, $\Theta^{(n)}$, $\Gamma^{(n)}$& Choi representation of $n$-slot quantum combs\\
        \hline
       $\mathsf{N}$, $\mathsf{T}$, $\mathsf{Q}$, $\mathsf{S}$ &  quantum testers\\
        \hline
       $N$, $T$, $Q$, $S$ & process effects (tester operators)\\
        \hline
    \end{tabular}
    \caption{Table depicts the association of symbols with the specific quantum objects.}
    \label{tab:my_label}
\end{table}
}
\subsection{Quantum combs}
\begin{figure}[h]
    \centering
    \includegraphics[width=0.45\textwidth]{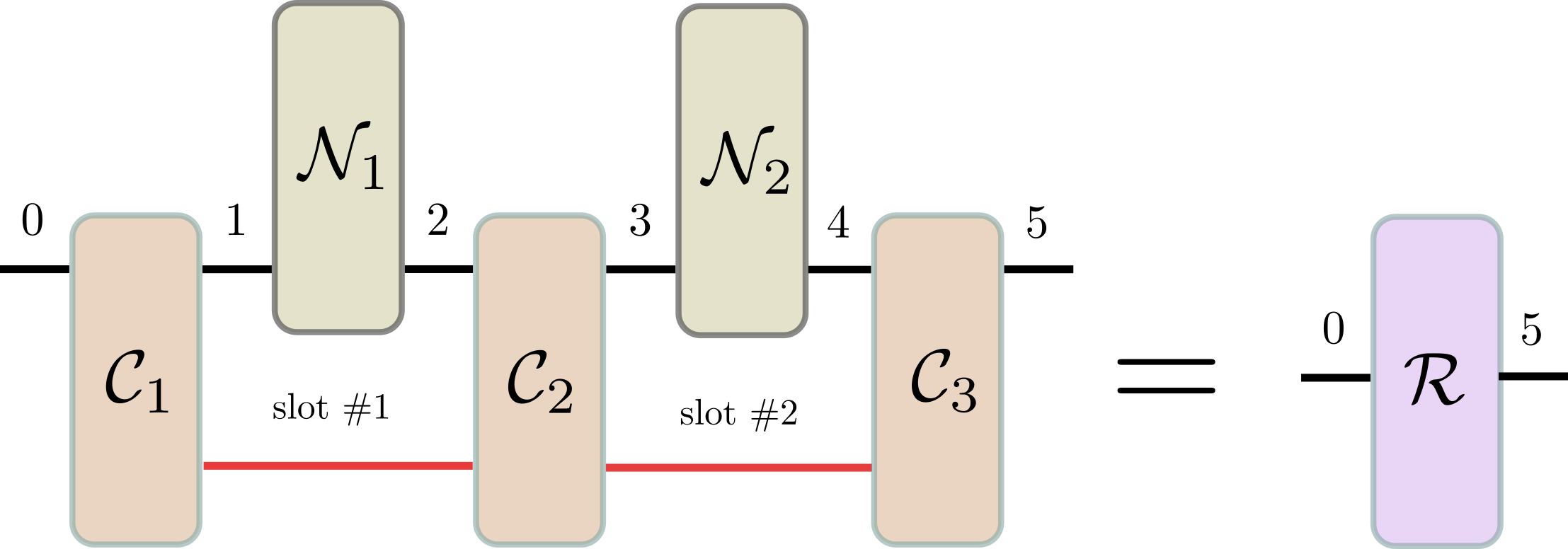}
    \caption{A 2-slot quantum comb (comprised of channels $\mathcal{C}_1$, $\mathcal{C}_2$, $\mathcal{C}_3$; Note that the red/light greyed out wires are quantum memories) transforming the two channels, $\mathcal{N}_1$ and $\mathcal{N}_2$, to a channel $\mathcal{R}$. In fact, the two slots can be plugged in with 1-slot combs. Here, $\mathcal{N}_1$ and $\mathcal{N}_2$ can be seen as a single 1-slot comb without any shared memory. }
    \label{fig:mesh1}
\end{figure}

\noindent Quantum channels, modeled by {completely-positive trace-preserving maps}, describe all quantum state transformations admitted by quantum theory. We can then generalise this notion of transformations by replacing states with quantum processes. These input and output processes can be channels, finite collections of time-ordered channels, superchannels or, most generally and equivalently, memory channels \cite{kretschmann2005quantum}. The {said} transformations are described by quantum circuits with open slots. Such incomplete circuits are described by \textit{quantum causal networks} or \textit{quantum combs} \cite{chiribella2008quantum,gutoski_quantumstrategies, chiribella2009theoretical}, where the number of open slots characterise the combs. For example, see Figure \ref{fig:mesh1}, for an illustration for the action of a 2-slot comb. It has also been identified that a $n$-slot quantum comb are equivalent to $(n+1)$ channels with memory \cite{chiribella2008memory}. It should also be noted that a comb describes all physical network implementations which result in identical transformations.

 As reflected in the figure, following the Pavia indexing convention \cite{chiribella2009theoretical}, for each $n$-slot comb, the input systems are indexed even $2k$ and output systems indexed odd $(2k+1)$, with $k \in \{0, \dots, n \}$. Now, we can write down the characterisation for  an $n$-slot quantum comb $\mathcal{C}^{(n)}$, with its Choi-Jamio\l kowski operator $\mathsf{C}^{(n)}$. The following linear constraints 
 on the Choi-Jamio\l kowski operator reflects the ``{causal structure}" of the comb, as well as the fact that the positive operators $\mathsf{C}^{(k)}$, for all values of $k$, are themselves Choi-Jamio\l kowski operators of combs with lesser and lesser number of open slots, on respective spaces.
\begin{eqnarray}\label{Eq:first}
   { \text{Tr}_{(2k+1)}[\mathsf{C}^{(k)}] = \mathds{1}_{(2k)} \otimes \mathsf{C}^{(k-1)} \quad; \ \forall k \in \{0, \dots, n \}.}
\end{eqnarray}
\noindent Now, from Figure \ref{fig:mesh1}, we 
 can notice that a 2-slot comb can transform a 1-slot comb to a channel (even though not depicted, it can also transform a 0-slot comb to a 1-slot comb). In general, we can see that an $n$-slot comb can transform $(k-1)$-slot combs to appropriate $(n-k-1)$-slot combs. These transformations generate hierarchical orders of transformations, referred to as \textit{higher-order transformations} \cite{chiribella2009theoretical}, enabling us to refer to this framework as higher-order quantum theory.
  
\subsection{Quantum testers}

\noindent The notion of quantum testers, introduced in \cite{chiribella2008memory,gutoski_quantumstrategies,ziman2008process}, describe measurements of quantum combs. That is, they describe measurement networks which send quantum combs to probability distributions. It should be noted that different physical implementations of measurement networks which result in identical probability distributions are described by the same quantum tester.\par
\begin{figure}[h]
    \centering
    \includegraphics[width=0.45\textwidth]{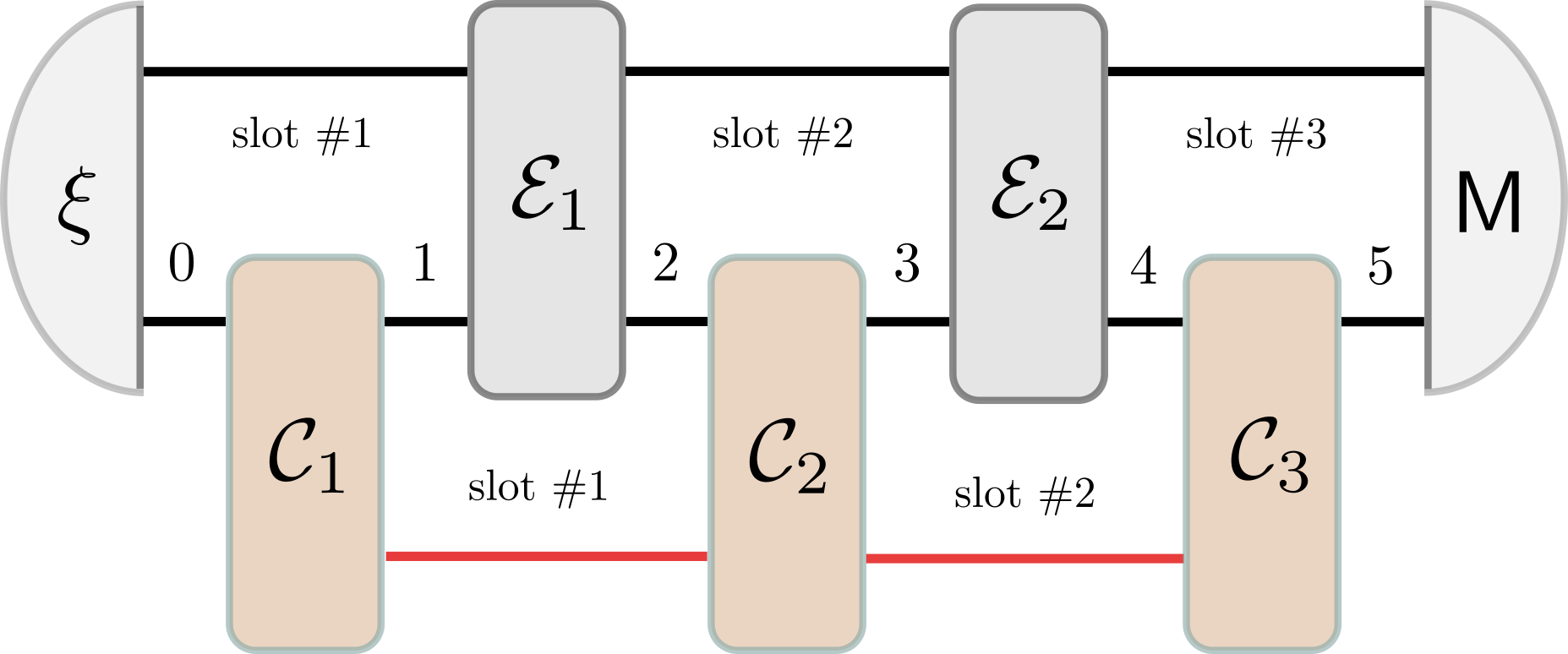}
    \caption{A 3-slot quantum tester (comprised of probe state $\xi$, channels $\mathcal{E}_1$, $\mathcal{E}_2$,  and probe POVM $\mathsf{M}$) measuring the 2-slot comb depicted in Figure 1.}
    \label{fig:mesh2}
\end{figure}

Similar to combs, testers are also characterised by the number of open slots they have. See Figure \ref{fig:mesh2}, for the action of a 3-slot tester. An $n$-slot tester measures $(n-1)$-slot combs. {As such,
we can see that 1-slot testers, which are nothing but single time-step quantum channels, test 0-slot combs.} Now, given  {a finite and discrete set of measurement outcomes $\Omega$ with $|{\Omega}| = \omega$, an $n$-slot tester $\mathsf{T}^{(n)}$ is a collection of positive operators referred to as ``process effects", $ {T}_1^{(n)}, \dots, {T}_{\omega}^{(n)},$ which satisfies the following {``normalisation conditions}" \cite{chiribella2008memory}. Sharing similarity with combs, these conditions reflect the causal structure of the network as well as the mathematical validity of the associated quantum components.}
{\begin{eqnarray}
    \sum_{x \in \Omega} {T}_x^{(n)} &=& \mathds{1}_{(2n-1)} \otimes \Xi^{(n)} \\
    \text{Tr}_{(2k-2)}[\Xi^{(n)}] &=& \mathds{1}_{(2k-3)} \otimes \Xi^{(k-1)} \quad \forall k \in \{2, \dots, n \} \\
    \text{Tr}[\Xi^{(1)}] &=& 1.
\end{eqnarray}}
{The operators, $\Xi^{(n)}, \dots, \Xi^{(2)}$, appearing above  are referred to as the ``{normalization objects}" associated with the tester $\mathsf{T}^{(n)}$, and are themselves quantum combs. As a remark on notations, the superscripts $(n)$, appearing on combs, testers and process effects, are dropped when specification of the number of associated open slots is irrelevant for the context.  }

\noindent { While implementing a tester $\mathsf{T} \equiv \{{T}_1, \dots, {T}_{\omega} \}$ on a given  comb} $\mathcal{C}$ with Choi operator $\mathsf{C}$, {the probability of observing an outcome labeled $k$ is furnished by the following generalised Born rule,}
\begin{equation}
    p(k| \mathcal{C}) = \text{Tr}[{T}_k \mathsf{C}].
\end{equation}
This Born rule, using the normalization of the tester ${\sum_x {T}_x = \mathds{1}_{(2n-1)} \otimes \Xi}$, can always be uniquely rewritten as to having the pair of a quantum state and a positive operator-valued measure (POVM) inside the trace functional. This POVM, which is  {uniquely prescribed by the given} tester, is referred to as its ``{canonical POVM}" \cite{chiribella2008memory}. It is given by
\begin{equation}
    {P}_j = {(\mathds{1}_{(2n-1)} \otimes \Xi^{-\frac{1}{2}})[{T}_j](\mathds{1}_{(2n-1)} \otimes \Xi^{-\frac{1}{2}})}.
\end{equation}
{For the rest of the work, when we refer to collections of testers (for example, $\{ \mathsf{T}_{\alpha}\}_{\alpha \in I}$  for some index set $I$) or to classical mixing of any given testers, the involved testers are assumed to have same number of open slots as well as identical sequences of input and output quantum systems; outcome sets can always set to be identical. Let $\mathsf{T} \equiv \{T_1, \dots, T_{\omega} \}$ and $\mathsf{S} \equiv \{ S_1, \dots, S_{\omega}\}$ be mixable testers. With $p \in [0,1]$, their mixture denoted as $p\mathsf{T} + (1-p)\mathsf{S}$ is described by the process effects, $pT_1 + (1-p)S_1, \dots, pT_{\omega} + (1-p)S_{\omega} $.} 

\subsection{Incompatibility of testers}
\noindent Similar to  how {a collection of} compatible POVMs {is} jointly measurable, {a compatible collection} of testers is jointly implementable. That is, there should exist {at least one joint or parent tester } {from which each of the testers, from the collection,} can be arrived at {through appropriate classical post-processings \cite{martens1990nonideal,heinonen2005optimal}}. {If there does not exist such a parent, then the collection is said to be incompatible.} Let $ \mathsf{T}^{(n)} $ and $\mathsf{S}^{(n)}$ be two $n$-slot testers, respectively, with normalisation {objects} $\{ \Xi^{(k)}\}$ and $\{ \Theta^{(k)}\}$. Then, {the pair of testers}  {is} compatible when their corresponding normalisations coincide, $\Xi^{(k)} = \Theta^{(k)}$ for all $k$, and their canonical POVMs are compatible themselves \cite{sedlak2016incompatible}. {These conditions for} incompatibility  {apply to any} finite collections of $n$-slot testers. 
 {Moreover, as it should be evident by now, given a tester $\mathsf{T}$, one can process it using different post-processings to result in a  collection of compatible testers.}  {Similar to how post-processing of a certain observable leads to a collection of compatible observables, post-processing of a fixed tester result in a collection of compatible testers, since the normalisation objects stay invariant and identical and, the canonical observables are post-processings of the parent one \cite{teikoIncWit}}.

 Now, given an incompatible collection of  testers 
 $\{ \mathsf{T}_{\alpha} \}$, {it can be rendered compatible by classically}  {mixing each tester with appropriate ``noise testers"} $\{ \mathsf{N}_{\alpha} \}$. {Here, the noise testers are restricted to be non-identical to each other (see subsection IV.C of \cite{sedlak2016incompatible}).} This  notion {of classical mixing of noise so as to facilitate ``breaking" of incompatibility serves as the basis for ``robustness of incompatibility", which was initially developed} for collections of {POVMs} \cite{RobustnessBusch, RobustnessTeiko, RobustnessHaapsaalo}.  Based on this notion and inspired from the one introduced in \cite{RobustnessUola+ConicProgramm}, we have {robustness of incompatibility} {$\mathcal{R}$} for the given collection of testers as, 
\begin{equation}
    \mathcal{R}(\{\mathsf{T}_{\alpha}\}) = \text{minimise } r : \left\{\frac{1}{1+r} (\mathsf{T}_{\alpha} + r\mathsf{N}_{\alpha}) \right\} \text{ is compatible.}
\end{equation}
{Similar to the robustness of  {incompatibility} of POVMs,  $\mathcal{R}$ possesses  {the following} desirable properties} \cite{sedlak2016incompatible}. {It is ``faithful" in the sense that, for a given $\{ \mathsf{T}_{\alpha}\}$, $\mathcal{R}(\{\mathsf{T}_{\alpha}\}) = 0$ if and only if the collection is compatible and non-zero otherwise. Moreover, it is monotonic under tester simulation  (mixing and post-processing) \cite{simulation_paper}. If a collection of testers  $\{\mathsf{T}'_{\beta}\}_\beta$ is simulated from another $\{ \mathsf{T}_{\alpha} \}$, then, 
  \begin{align}
        \mathcal{R}(\{\mathsf{T}'_{\beta}\})\leq \mathcal{R}(\{\mathsf{T}_{\alpha}\}).
    \end{align}
    One such simulation that we shall encounter later in our investigations is given by
    \begin{align}
    \mathsf{T}'_{b|\beta}=\sum_{a,\lambda,\alpha} p(\lambda)p(\alpha|\beta,\lambda)p(b|a,\beta,\lambda)\mathsf{T}_{a|\alpha}.
    \end{align}
    For the moment, it is adequate enough to consider the $p(\cdot)$'s appearing above to be bona fide probability distributions over the associated variables.}

\section{Advantage in quantum comb discrimination game (QCD)}\label{III1}
\subsection{QCD} 
\noindent {We denote an ensemble of $m (\in \mathbb{N})$ $n$-slot quantum combs as $\mathcal{X} = \{ (\mathcal{C}_b, w(b))\}_{b=1}^m$, where $w(b)$ is the probability with which the comb $\mathcal{C}_b$ is chosen from the ensemble. One can have finite collections of such ensembles $\{\mathcal{X}_{\beta}\}_{\beta = 1}^s$ for some $s\in \mathbb{N}$. Now, a specific ensemble $\mathcal{X}_{\beta}$ can be represented as $\mathcal{X}_{\beta} = \{ (\mathcal{C}_{b|\beta}, w(b|\beta))\}_{b=1}^{m}$. Here, $\mathcal{C}_{b|\beta}$ is the $b^{\text{th}}$ comb of the $\beta^{\text{th}}$ ensemble, chosen with conditional probability $w(b|\beta)$.
The two party QCD game and associated notions, with Bob being the referee and Alice the player, is sketched as follows.}
{
\begin{enumerate}
    \item Bob has access to a collection of ensembles $\{\mathcal{X}_{\beta}\}_{\beta = 1}^s$. Alice, in prior to playing the game, is aware of the information regarding these ensembles. On the other hand, Alice has access to a  fixed, finite  collection of testers $\{ \mathsf{T}_{\alpha} \}_{\alpha=1}$. Each tester $\mathsf{T}_{\alpha}$ is composed of $m$ process effects, $T_{1|\alpha}, \dots, T_{m|\alpha}$, where $T_{a|\alpha}$ corresponds to the measurement event of guessing $a^{\text{th}}$ comb from any given ensemble. 
    \item In each round of the game, he chooses an ensemble $\mathcal{X}_{\beta}$ out of this collection with probability $w(\beta)$. From the chosen ensemble he picks out a comb $\mathcal{C}_{b|\beta}$ with probability $w(b|\beta)$.
    \item He sends Alice a black box that implements the chosen comb $\mathcal{C}_{b|\beta}$, along with the classical side information $\beta$ associated with his choice.
    \item Alice, now equipped with the knowledge regarding which ensemble was chosen by Bob, is now tasked to identify the acquired comb, or equivalently guess the value of $b$. She plays this guessing game by performing a  distinguishability experiment using testers she has access to.  
    \item  Suppose Alice picks up $\alpha^{\text{th}}$ tester $\mathsf{T}_{\alpha}$, conditioned on availed information and/or using classical randomness accessible to her, and ``couples" it with the acquired comb, resulting in a measurement outcome $a$. Correspondingly, her guess is the $a^{\text{th}}$ comb from the $\beta^{\text{th}}$ ensemble, i.e., $\mathcal{C}_{b|\beta}$. The probability of this guess is furnished by $p(a|b)={\rm Tr}[\mathsf{C}_{b|\beta} {T}_{a|\alpha}]$.
\end{enumerate}
}
{With the game sketched out, we present the main result in the subsequent subsection.}


\subsection{ Tester incompatibility as a resource in QCD}

\noindent {To elucidate the resourcefulness of incompatibility, we consider the following two scenarios of game-play with which Alice can play the QCD game.} 

\vspace{0.5cm}

\noindent {\it \textbf{Scenario 1: {Alice has access to a collection of incompatible testers}}}.-- In this gameplay, Alice's access is restricted to incompatible collections of testers $\{ \mathsf{T}_{\alpha}\}_\alpha$. After receiving the value of {side-information} $\beta$ and the black box, by having access to a {classical} random variable $\lambda$, Alice chooses and implements the tester $\mathsf{T}_{\alpha}$ with probability $p(\alpha | \beta, \lambda)$. From the measurement outcome $a$, she guesses the value of $b$ as prescribed by $p(g|a,\beta,\lambda)$. {As such, the different strategies she can adopt are determined by the triple of  probability distributions, $p(\lambda), p(\alpha|\beta,\lambda)$ and $p(g|a,\beta,\lambda)$}. She can then evaluate the {optimal} average probability of correct guessing $g$
by optimizing over all possible strategies, { $\mathcal{S}=\{ ( p(\lambda), p(\alpha|\beta,\lambda),p(g|a,\beta,\lambda) )\}  $}, 

\begin{eqnarray}
	p(\{\mathcal{X}_{\beta}\},\{ \mathsf{T}_{\alpha}\}) = { \max_{\mathcal{S}} \sum_{b,\beta,a,\alpha,g,\lambda} w(b,\beta)p(\lambda) p(\alpha|\beta,\lambda)  \text{Tr} [\mathsf{C}_{b|\beta}{T}_{a|\alpha}]  p(g|a,\beta,\lambda) \delta_{g,b}.}
\end{eqnarray}
{Here, the summands are probabilities of all possible successful guesses; this is enforced by the delta functions $\delta_{g,b}$ and $w(b,\beta)$ is identified with $w(\beta)w(b|\beta)$. }

\vspace{0.5cm}

\noindent \textbf{\textit{Scenario 2: {Alice has access to  collections of compatible testers}}}.-- {In the aforementioned scenario, Alice also uses the side-information to help her choose the tester, which she then implements. In this second scenario, within each run, her choice of tester is not prescribed by the side-information but rather due to a classical random variable $\nu$ alone. Thus, in each run, the tester $\mathsf{Q}_{\nu}\equiv \{ Q_{1|\nu}, \dots, Q_{m|\nu} \}$ is fixed with probability $p(\nu)$. After implementing this tester, she can only use the side-information to post-process the measurement outcome or, equivalently, the fixed tester $\mathsf{Q}_{\nu}$. Thus, essentially she is implements, or equivalently has access to, compatible collections of testers. Each such collection has its parent tester as one of $\{ \mathsf{Q}_{\nu}\}$. As such, the different strategies she can adopt are determined by the probability $p(\nu)$, the parents $\mathsf{Q}_{\nu}$ and the guessing conditional probabilities $p(g|a, \beta, \nu)$. Thus, the optimal guessing success probability is given by,}  
\begin{eqnarray}
    p(\{\mathcal{X}_{\beta}\}) = \max_{\mathcal{S}_c} \sum_{b,\beta,a,g,\nu}  w(b,\beta)p(\nu)  \text{Tr}[\mathsf{C}_{b|\beta}\mathsf{Q}_{a|\nu}]p(g|a,\beta,\nu)\delta_{g,b}.
\end{eqnarray}
\vspace{0.5cm}
Here, optimization is done over all strategies, $\mathcal{S}_c=\{p(\nu),\mathsf{Q}_{\nu},p(g|a,\beta,\nu)\}$. 

When maximised over all possible 
games, the maximum value of the relative increase in the guessing probability attainable by  {any} given set of incompatible testers $\{ \mathsf{T}_{\alpha}\}_\alpha$ with respect to that is attainable only by single parent testers  is given by
\begin{equation}
	\max_{\{\mathcal{X}_{\beta}\}} \frac{  p(\{\mathcal{X}_{\beta}\},\{ \mathsf{T}_{\alpha}\})}{  p(\{\mathcal{X}_{\beta}\})}.
\end{equation}
We find that this maximum ``advantage" attainable by Alice  is related to the robustness of incompatibility  and is encapsulated into the following theorem.

\begin{theorem} 
For the QCD game, while maximised over all possible game ensembles, the relative performance of having access to any collection of incompatible testers $\{ \mathsf{T}_{\alpha} \}$, and following game-play scenario 1, to having access to compatible collections through virtue of following game-play scenario 2, is always greater than 1. This advantage is captured by,
\begin{align}\label{desired}
   \max_{\{\mathcal{X}_{\beta}\}} \frac{  p(\{\mathcal{X}_{\beta}\},\{ \mathsf{T}_{\alpha}\})}{  p(\{\mathcal{X}_{\beta}\})}=1+ \mathcal{R}(\{\mathsf{T}_{\alpha}\}).
\end{align}
\end{theorem}

\noindent {We refer enthusiastic readers to Appendix \ref{appen:1} for a detailed proof of the theorem using semi-definite programming arguments. Our proof follows closely to that of in  Ref. \cite{Skrzypczyk2019}; however our investigation generalizes their findings to higher-order incompatibility and associated potential ramifications are discussed in the conclusions. }  

Now,   conclude that the advantage using incompatible testers in our QCD game is non-increasing under tester simulation discussed above. This can be understood in following way: Under tester simulation, one will not be able to elevate the compatible set of testers to incompatible ones. Moreover, under such action, the robustness of tester incompatibility decreases. Therefore, {one would expect} the gain in QCD game will also monotonically decrease under tester simulation. For completeness, we refer our readers to the Ref. \cite{Skrzypczyk2019} where it is discussed in detail.

\section{The ``convex weight" for quantum testers, and advantages in exclusion games of comb}\label{IV1}
\textcolor{black}{In \cite{uola2020all}, the authors have found that, using the quantum resource theoretic measure \textit{convex weight} and auxiliary mathematical programming arguments, any quantum resource improves performance of quantum exclusion tasks/games of zeroth order quantum devices, including states, POVMs, channels and instruments. These  {findings have}  been bolstered by the subsequent work \cite{PhysRevLett.125.110401}. In general, since the result hold{s} for quantum instruments, it can be recast for the rest of the devices mentioned  {alongside}. Since higher-order quantum devices also admit equivalent analyses with minor changes, we would expect that for exclusion games of higher-order quantum devices, any resource would result in improvement. This suggests that incompatibility is a resource, not just for distinguishability games but also for exclusion or anti-distinguishability games \cite{bandyopadhyay2014conclusive, caves2002conditions}.  }

Convex weight for an arbitrary quantum device{s} is defined in Ref. \cite{uola2020all}.
The convex weight fulfills all the desirable properties to be a useful resource quantifiers similar to robustness of quantum devices. {To keep everything simple and easy reading for the readers, we will define convex weight of incompatibility for quantum testers. Note here that the convex weight can be defined for any resource present in the tester by changing the set of free testers. As set of collections of compatible testers form a convex and compact set and is the ``free" set when considering incompatibility as a resource, let us denote it as $ \mathcal{T}_F$. Then, for incompatible testers,}  $\{ \mathsf{T}_{\alpha} \}$ with $ \mathsf{T}_{\alpha}\equiv \{T_{a|\alpha}\}_a$, we can define its convex weight  as {``the maximum relative number of times a collection of testers from the set $\mathcal{T}_F$ is used to produce the collection of testers $\{ \mathsf{T}_{\alpha} \}$", i.e.,} 
\begin{align}\label{conic0}
    \mathcal{W}(\{\mathsf{T}_{\alpha}\}):&=\min s_w\\
    &{\rm s.t.}\quad T_{a|\alpha}=(1-s_w) Q_{a|\alpha}+s_w \Bar{T}_{a|\alpha},\nonumber
\end{align}
{where $\{\Bar{\mathsf{T}}_\alpha\}$, with $\Bar{\mathsf{T}}_\alpha\equiv \{\Bar{T}_{a|\alpha}\}_a$, are the collections of general testers outside the set $\mathcal{T}_F$ and the collection $\{\mathsf{Q}_\alpha\}\in \mathcal{T}_F$ where $\mathsf{Q}_\alpha\equiv \{Q_{a|\alpha}\}_a$. By defining $\Bar{Q}_{a|\alpha}=(1-s_w) Q_{a|\alpha}$ and exploiting the property of Choi-Jamio\l kowski representation, we notice that $(1/s_w)(T_{a|\alpha}-\Bar{Q}_{a|\alpha})$ must be positive $\forall$ $(a,\alpha)$. Notice here that the set of collections of testers $\{\Bar{\mathsf{Q}}_\alpha\}$ with $ \Bar{\mathsf{Q}}_\alpha\equiv \{\Bar{Q}_{a|\alpha}\}$ forms a cone, i.e., $\{\Bar{\mathsf{Q}}_\alpha\}\in \mathsf{Cone}_F$, where $\mathsf{Cone}_F:=\{\gamma \{\mathsf{Q}_\alpha\}|\gamma>0, \{\mathsf{Q}_\alpha\}\in\mathcal{T}_F\}$ is a cone based on the subset $\mathcal{T}_F$. Also, noticing that $1-s_w=(1/\alpha D){\rm Tr}[\sum_{a,\alpha}\Bar{Q}_{a|\alpha}]$, we reach to the following coning programming with linear constraints}
\begin{align}\label{conic1}
    1-\mathcal{W}(\{\mathsf{T}_\alpha\})&=\max \frac{1}{n D}{\rm Tr}\left[\sum_{a,\alpha}\Bar{Q}_{a|\alpha}\right]\\
    &{\rm s.t.}\quad T_{a|\alpha}\geq \Bar{Q}_{a|\alpha} \quad\forall (a,\alpha),\nonumber\\
&\quad\quad\quad\{\Bar{\mathsf{Q}}_\alpha\}\in \mathsf{Cone}_F,\nonumber
\end{align}
where {$n$ is the total number of testers in any given collection of testers, i.e., the largest value $\alpha$ takes as an index variable and} $D = \prod_{j=0}^{n}d_{2n-(2j+1)}$ where $d_x$ is the dimension of the system indexed $x$ and $\alpha$ is the number of testers. The dual formulation of the above conic program always exists \cite{Skrzypczykbook} and it reads as below
\begin{align}\label{conic2}
    1-\mathcal{W}(\{\mathsf{T}_\alpha\})&=\min_{Y_{a|\alpha}} {\rm Tr}\left[\sum_{a,\alpha}T_{a|\alpha} Y_{a|\alpha} \right],\\
   & {\rm s.t.}\quad Y_{a|\alpha}\geq 0 \quad\forall (a,\alpha),\nonumber\\
    &{\rm Tr}\left[\sum_{a,\alpha}Q_{a|\alpha} Y_{a|\alpha} \right]\geq 1,\quad \{\mathsf{Q}_\alpha\}\in \mathcal{T}_F,\nonumber
\end{align}
{where $\mathsf{Y}:=\bigoplus_{a,\alpha} Y_{a|\alpha}$ is a witness \cite{uola2020all}. If we choose $\mathsf{Y}=\gamma \id$ for $\gamma>0$, then the solution to primal and dual merges implying the verification of Slater condition of strong duality.} 

{{\bf \textit{Comb exclusion game}}.-- Referee Bob has access to the collection of ensembles $\{\mathcal{X}_{\beta}\}_{\beta}$. During each run of the game, Bob chooses an ensemble $\mathcal{X}_{\beta}$ and then a comb $\mathsf{C}_{b|\beta}$ from the ensemble with probability $w(b,\beta)$. Then, he sends a black box implementing the comb and the associated  side-information $\beta$ to player Alice. Alice succeeds at the game if she guesses a comb which is not sent to her or excludes the acquired comb or, equivalently, $b$.}

Given the structure of the game, {if Alice can only choose the testers $\{\mathsf{Q}_{\alpha}\}$} from  $\mathcal{T}_F$, the minimum {averaged} error probability {of excluding $b$, given the side-information $\beta$,} is given by 
\begin{align}
    \Bar{p}(\mathcal{X}_{\beta})=\min_{{\{\mathsf{Q}_{\alpha}\}}\in\mathcal{T}_F}\sum_{a,b} w(b,\beta){\rm Tr}\left[Q_{a|\alpha} \mathsf{C}_{b|\beta}\right].
\end{align}
{Notice in the above equation that the success probability of excluding a comb is exactly equal to $1-\Bar{p}(\mathcal{X}_{\beta})$.}
Can Alice do better if she has access to set of incompatible testers $\{\mathsf{T}_\alpha\}_\alpha$? {As the referee is communicating the value of $\beta$, Alice will always choose the set of testers of the form $\mathsf{T}_\alpha$ with outcome set $\alpha$ which will have least overlap with the communicated set $\beta$. Then, the minimum-error probability of successfully excluding the  comb is given by,}
\begin{align}
    \Bar{p}(\mathcal{X}_{\beta},\mathsf{T}_\alpha)=\min_{\mathsf{T}_\alpha}\sum_{a,b}w(b,\beta){\rm Tr}\left[T_{a|\alpha} \mathsf{C}_{b|\beta}\right].
\end{align}
{Again, it is evident in the above equation that $1-\Bar{p}(\mathcal{X}_{\beta},\mathsf{T}_\alpha)$ is the success probability of excluding the comb.} 
In general, the player will have less error with the strategies involving incompatible testers than its compatible counterparts, i.e., $\Bar{p}(\mathcal{X}_{\beta},\mathsf{T}_\alpha)/\Bar{p}(\mathcal{X}_{\beta})\leq 1$ \cite{uola2020all}. In fact, we are able to prove the following result, using the similar techniques deployed in the Ref. \cite{uola2020all} (see also Appendix \ref{conv-app} for an alternative proof),
\begin{theorem}
    {In any comb exclusion game, having access to incompatible testers are always advantageous over the compatible ones. Moreover, the relative advantage is exactly quantified by the following relation,}
    \begin{align}
    \min_{\mathcal{X}_{\beta}} \frac{\Bar{p}(\mathcal{X}_{\beta},\mathsf{T}_\alpha)}{\Bar{p}(\mathcal{X}_{\beta})}=1-\mathcal{W}(\{\mathsf{T}_\alpha\}).\label{conv-res}
\end{align}
\end{theorem}
{
\begin{proof}
    Now from dual conic program, we define a new variable 
    \begin{align}
        Y^*_{a|\alpha}=Y_{a|\alpha}\Big/q^*(a,\alpha)N, \quad {\rm s.t.} \quad {\rm Tr}[Y^*_{a|\alpha}]=\Bar{D} \quad \forall (a,\alpha), \quad {\rm and} \quad q^*(a,\alpha)={\rm Tr}[Y_{a|\alpha}]\Big/ \sum_{a,\alpha}{\rm Tr}[Y_{a|\alpha}],
    \end{align}
    which implies that $N=(1/\Bar{D})\sum_{a,\alpha}{\rm Tr}[Y_{a|\alpha}]$ and $\{q^*(a,\alpha)\}$ is a normalized probability distribution. To ensure that $Y^*_{a|\alpha}$ are combs, $\Bar{D}$ should depict the total dimension of the output space of the comb, i.e., $\Bar{D} = \prod_{j=0}^{n}d_{2n-(2j+1)}$ with $d_x$ is the dimension of the system labeled ``$x$".
    Now using the new witness and  Eq. \eqref{conic0}, we find that 
    \begin{align}
        \Bar{p}(\mathcal{X}_{\beta},\mathsf{T}_\alpha)\geq [1-\mathcal{W}(\{\mathsf{T}_\alpha\})]\min_{\mathsf{Q}_\alpha\in\mathcal{T}_F} \Bar{p}(\mathcal{X}_{\beta}, \mathsf{Q}_\alpha),
    \end{align}
    where $\min_{\{\mathsf{Q}_\alpha\}\in\mathcal{T}_F} \Bar{p}(\mathcal{X}_{\beta}, \mathsf{Q}_\alpha):=\Bar{p}(\mathcal{X}_{\beta})$. 
    Subsequently, using Eq. \eqref{conic2}, and noting that the scaling used in our definition of $Y^*$ will not effect to the overall game, we get following relation
    \begin{align}
        \min_{\mathcal{X}_{\beta}} \frac{\Bar{p}(\mathcal{X}_{\beta},\mathsf{T}_\alpha)}{\Bar{p}(\mathcal{X}_{\beta})}=1-\mathcal{W}(\{\mathsf{T}_\alpha\}).
    \end{align}
    Note here that we choose a suitable ensemble $\mathcal{X}_{\beta}$ for which the above optimization is always finite.
\end{proof}
}
The above result in Eq.\eqref{conv-res} implies that there exists a strategy where the decrease in the minimum-error probability in a comb exclusion game is given by the convex weight of incompatibility in the quantum device available to the player. {By noticing the generality of above formulation, we find that the following general result also holds:
\begin{corollary}
    For any resource present in the tester, if the resource free testers forms a convex and compact set $\mathcal{T}_F$, then any testers $\mathsf{T}\notin \mathcal{T}_F$ will yield advantages in any minimum-error exclusion game of quantum combs. Moreover, the relative advantage will be  quantified by the convex weight of that resource.
\end{corollary} 
}
\section{Conclusion}\label{V1}
\noindent The notion of quantum testers was introduced to describe the testing or measurements of multiple time-step quantum causal networks, now popularised as quantum combs. Since these testers are generalised quantum measurements, the operational notion of quantum incompatibility also has been introduced and developed for them, where we study collections of testers  {which} seldom admit joint testers. In the wake of all of these developments, in this work, we investigated the quality of resourcefulness of incompatible testers. Consequently, we showed that the incompatible testers result in advantages, over compatible ones, in a specific game in which quantum combs are distinguished. This heralds that the incompatibility, as a quantum feature, present in general test procedures is a resource in discriminating higher-order quantum objects. Moreover, this generalises the earlier finding that `POVM incompatibility is a resource for distinguishability of states', which was discovered in Ref.\cite{Skrzypczyk2019}. One of the main differences from our work is that the aforementioned work was based on the discrimination of quantum objects of static nature, namely, quantum states to demonstrate the resourcefulness of incompatible quantum test procedures for states; In our work, where central objects of interest are quantum combs, they describe networks composed of quantum objects of both nature: static as well as dynamical. In this context, the result of \cite{Skrzypczyk2019}, as qualitatively powerful as it is, can be viewed as a special case of our generalisation when the associated combs are restricted to be states. Another peculiarity of our result is that it encompasses all distinguishability or discrimination schemes possible for any basic quantum object (including states, single time-step operations and POVMs): irrespective of whether it is single-shot or multiple-shot and whether it is parallel or adaptive. Moreover, it encompasses distinguishability of quantum channels with memory  channels\cite{kretschmann2005quantum}.

Moreover, we investigated quantum comb exclusion games and discovered that any quantum resource present in quantum testers result in performance improvement in such games. Our result generalizes the result found in \cite{uola2020all} which dealt with resourcefulness of POVMs in state exclusion games. Quantum resources of general testers are contributed from the component quantum objects of the network as well as potentially from the quantum memories they possess. As a consequence of our result, quantum incompatibility of testers, which arise in the same way, can then be identified as a resource in quantum comb exclusion games. In conclusion, we have discovered that the more intricate notion of higher-order incompatibility is a resource in quantum hypothesis-testing tasks involving combs, including distinguishability and exclusion.


\begin{acknowledgments}
\noindent NSR acknowledges  the support of projects APVV-22-0570 (DeQHOST) and VEGA 2/0183/21 (DESCOM). NSR also acknowledges Michal Sedl{\'a}k for clarifying certain concepts regarding quantum combs, {and Anna Jenčová for drawing attention to technical errata in initial draft of the work.} SS acknowledges funding through PASIFIC program call 2 (Agreement No. PAN.BFB.S.BDN.460.022 with the Polish Academy of Sciences). This project has received funding from the European Union’s Horizon 2020 research
and innovation programme under the Marie Skłodowska-Curie grant agreement No 847639 and from the Ministry of Education and Science of Poland. {We acknowledge the anonymous referees for their valuable feedback that improved the clarity and readability of the work.}
\end{acknowledgments}

\section*{Author declarations}
\noindent The authors have no conflicts to disclose.

\section*{Data Availability Statement}
\noindent Data sharing is not applicable to this article as no new data were created or analyzed in this study.

\bibliography{tester}
\onecolumngrid
\newpage
\appendix

\section{Tester incompatibility results in advantages in QCD}\label{appen:1}

\subsection{Noise robustness} \label{noise-primal}

\noindent As we have seen, Eqs(2-4) prescribe  $n$-slot testers. We consider a finite incompatible collection of such principal testers $\{ \mathsf{T}_{\alpha}\}_\alpha$ {with $\mathsf{T}_{\alpha}\equiv \{T_{a|\alpha}\}_a$ $\forall \alpha$}. The the robustness of this collection $\mathcal{R}(\{ \mathsf{T}_{\alpha}\})$ is defined as the follows, from which we can identify a compatible collection of testers, corresponding to parent observable $\mathsf{Q}\equiv \{Q_{\mu}\}_\mu$ and post-processing probability distribution $p(a|\alpha,\mu)$ as appearing below.
\begin{eqnarray}
\mathcal{R}(\{ \mathsf{T}_{\alpha} \}) &=& \min \:\:\: r \:\: \\
&& \text{s.t. }\frac{1}{(1+r)} (T_{a|\alpha} + r N_{a|\alpha}) = \sum_{\mu} p(a|\alpha,\mu) Q_{\mu}, \label{Eq:A2}\\
\text{[principal testers]}&& T_{a|\alpha} \ge 0, \sum_{a} T_{a|\alpha} = \mathds{1}_{(2n-1)} \otimes \Theta^{(n)}_{\alpha} ,\\
&& \text{Tr}_{(2k-2)}\{\Theta^{(k)}_{\alpha}\} =  \mathds{1}_{(2k-3)} \otimes \Theta^{(k-1)}_{\alpha} \quad \forall k \in [2,n], \\
&&\text{Tr}\{\Theta_{\alpha}^{(1)}\}=1. \\ \\
\text{[noise testers]}&& N_{a|\alpha} \ge 0, \sum_{a} N_{a|\alpha} = \mathds{1}_{(2n-1)} \otimes \Gamma^{(n)}_{\alpha}, \\
&& \text{Tr}_{(2k-2)}\{\Gamma^{(k)}_{\alpha}\} =  \mathds{1}_{(2k-3)} \otimes \Gamma^{(k-1)} \quad \forall k, \\
&&\text{Tr}\{\Gamma_{\alpha}^{(1)}\}=1.\\ \\ 
\text{[probability distribution]}&& p(a|\alpha,\mu) \ge 0, \sum_{\mu} p(a|\alpha,\mu) = 1. \\ 
\text{[joint tester]}&& Q_{\mu} \ge 0, \sum_{\mu} Q_{\mu} = \mathds{1} \otimes \Theta^{(n)}, \\
&& \text{Tr}_{(2k-2)}\{\Theta^{(k)}\} =  \mathds{1}_{(2k-3)} \otimes \Theta^{(k-1)} \quad \forall k, \\
&&\text{Tr}\{\Theta^{(1)}\}=1.\\ \\
\text{[normalisation objects]}&& \Theta^{(k)} = \frac{1}{(1+r)} (\Theta^{(k)}_{\alpha} + r\Gamma^{(k)}_{\alpha} ) \quad \forall \alpha, k, \label{Eq:a18}\\
&& \Theta^{(k)}_{\alpha} \ge 0 \; \text{ and } \; \Gamma^{(k)}_{\alpha} \ge 0 \quad \forall \alpha, k.\label{Eq:a19}
\end{eqnarray}
{The noise testers are denoted by $\{\mathsf{N}_\alpha\}_\alpha$ with $\mathsf{N}_\alpha\equiv \{N_{a|\alpha}\}_a$. All the tester normalisation objects are given by the $k$-slot combs, $\Theta^{(k)}_{\alpha}$, $\Gamma^{(k)}_{\alpha}$ and $\Theta^{(k)}$ where $k \in [2,n]$. The extra conditions on normalization objects given in Eqs. \eqref{Eq:a18}-\eqref{Eq:a19} are coming from the requirements for a set of testers to be compatible \cite{sedlak2016incompatible}. Notice also that in order to be a valid comb, the $\Theta^{(k)}_{\alpha}$, $\Gamma^{(k)}_{\alpha}$ and $\Theta^{(k)}$ should satisfy certain conditions dictated by Eq. \eqref{Eq:first}. Also, note that the above SDP is a primal one for our analysis. Below, we will show that the primal SDP can be simplified further using arguments below.}
\subsection{Primal SDP}
\noindent See Chapter 6 of \cite{Skrzypczykbook} on measurement incompatibility, as well as the work \cite{Skrzypczyk2019}, where the arguments that we use in the formulation of the following SDP can be found. From Eq. \eqref{Eq:A2}, the process effects of the noise testers are given by,
\begin{equation}
    N_{a|\alpha} = \frac{1}{r}\{ (1+r) \sum_{\mu}  p(a|\alpha,\mu) Q_{\mu} -T_{a|\alpha} \}.
\end{equation}
Then, with $s = (1+r)$, the positive semi-definiteness of the process effects results in,
\begin{equation}
    s \sum_{\mu}  p(a|\alpha,\mu) Q_{\mu} \ge T_{a|\alpha} .
\end{equation}
Now, let $a_k$ be the outcome label for the $k^{\text{th}}$ tester. As such,  we can collectively write all the $m$ outcome labels into a vector $\vec{a} = (a_1, \dots, a_m)$. Since, each of the $m$ testers has $o$ outcomes, that is $a_k \in \{1, \dots, o \}$ for all $k$,  there are $o^m$ possible vectors. Given the result $\mu$ of the joint tester, the probability of producing a specific vector $\vec{a}$ is given by $p(\vec{a}|\mu) = p(a_1 | \alpha =1, \mu)p(a_2 | \alpha =2, \mu)\cdots p(a_m| \alpha = m, \mu) = \prod_{\alpha} p(a_{\alpha} | \alpha, \mu)$. Now, the probability of the $\alpha^{\text{th}}$ entry being equal to $a$, that is $a_{\alpha} =a$, is given by $p(a|\alpha,\mu) = \sum_{\vec{a}} \delta_{a_{\alpha},a} p(\vec{a}|\mu)$.   Then, conditional probability distribution can be decomposed using the basis of deterministic conditional probabilities, $\{ d_{\Vec{a}}(a|\alpha) = \delta_{a,\vec{a}_x}\}$ as $p(a|\alpha,\mu) = \sum_{a}d_{\Vec{a}}(a|\alpha)p(\vec{a}|\mu)$, where 
\begin{equation}
    \sum_{\mu} p(a|\alpha,\mu) Q_{\mu} = \sum_{\vec{a}}d_{\Vec{a}}(a|\alpha) Q_{\vec{a}},
\end{equation}
while $Q_{\vec{a}} = \sum_{\mu} p(\vec{a}| \mu) Q_{\mu}$. Since $Q_{\vec{a}} \ge 0$, we can define $\tilde{Q}_{\vec{a}} := sQ_{\vec{a}}$. Then,
\begin{eqnarray}
&& \tilde{Q}_{\vec{a}} \ge0 \\
&& \sum_{\vec{a}}\tilde{Q}_{\vec{a}} = s \mathds{1} \otimes \Theta^{(n)} \\
&& \text{Tr}_{(2k-2)}\{\Theta^{(k)}\} =  \mathds{1}_{(2k-3)} \otimes \Theta^{(k-1)} ; \quad k \in \{ 2, \dots, n\}
\end{eqnarray}
This means we essentially find the parent tester which has the following normalisation (see Eq. \eqref{Eq:a18}),
\begin{eqnarray}
\Theta^{(k)} = \frac{1}{s} (\Theta^{(k)}_{\alpha} + r\Gamma^{(k)}_{\alpha} ) \quad \forall \alpha \: \forall k.
\end{eqnarray}
In fact, the above conditions are automatically satisfied since post-processings preserve normalisation and as such does not appear in the SDP. Therefore, we can re-formulate the above SDP (defined in Subsec. \ref{noise-primal}) in the following simple form,
\begin{eqnarray}
1 + \mathcal{R}(\{ \mathsf{T}_{\alpha} \}) &=& \min \:\:\:s \label{Eq:A26}\\
 && \text{s.t. } \sum_{\Vec{a}} d_{\Vec{a}}(a|\alpha) \Tilde{Q}_{\Vec{a}}\geq T_{a|\alpha} \\
 &&\tilde{Q}_{\vec{a}} \ge 0, \; \sum_{\vec{a}}  \tilde{Q}_{\vec{a}} = \mathds{1} \otimes \tilde{\Theta}^{(n)} \\
 && \text{Tr}_{(2k-2)}\{\tilde{\Theta}^{(k)}\} =  \mathds{1}_{(2k-3)} \otimes \tilde{\Theta}^{(k-1)} ;\:\: \forall \quad k \in \{2, \dots, n \},\label{Eq:A29}
\end{eqnarray}
{where we have defined the modified combs defined as $\tilde{\Theta}^{(n)}:=s\Theta^{(n)}$ using the liberty of Choi representation.}
\subsection{Lower bound}
\noindent We can use the primal SDP to have an upper bound on the advantage in the QCD game. In order to achieve this, let us start by assuming that one of the solutions to our SDP shall be achieved by the triple $(N_{a|\alpha}^*, Q_\mu^*,p^*(a|\alpha,\mu))$. Then, we have the following 
\begin{align}
    \frac{1}{1+\mathcal{R}(\{\mathsf{T}_{\alpha}\})}\left(\mathsf{T}_{a|\alpha}+\mathcal{R}(\{\mathsf{T}_{\alpha}\})N_{a|\alpha}^*\right)=\sum_\mu p^*(a|\alpha,\mu)Q_\mu^*.
\end{align}
Since $\mathcal{R}(\{\mathsf{T}_{\alpha}\})\geq 0$ and $N_{a|\alpha}^*\geq 0$, the above equation results in the inequality,
\begin{align}
    [1+\mathcal{R}(\{\mathsf{T}_{\alpha}\})]\sum_\mu p^*(a|\alpha,\mu)Q_\mu^*\geq T_{a|\alpha},\:\: \forall a,\alpha.
\end{align}
Now multiplying both sides of the inequality with $\mathsf{C}_{b|\beta}$, the probabilities $w(b,\beta), p(\nu),  p(\alpha|\beta,\nu),  p(g|a,\beta,\nu)$ and $ \delta_{g,b}$ and then summing over the corresponding indices and taking trace, we arrive at the inequality,
\begin{align}\label{Eq:12}
    [1+\mathcal{R}(\{\mathsf{T}_{\alpha}\})]\sum_{\mu,g,a,b,\alpha,\beta,\nu}w(b,\beta)p(\nu) p(\alpha|\beta,\nu) p^*(a|\alpha,\mu)\text{Tr} [\mathsf{C}_{b|\beta}Q_\mu^*]  p(g|a,\beta,\nu) \delta_{g,b}\nonumber\\\geq \sum_{\mu,g,a,b,\alpha,\beta,\nu}w(b,\beta)p(\nu) p(\alpha|\beta,\nu) \text{Tr} [\mathsf{C}_{b|\beta}T_{a|\alpha}]  p(g|a,\beta,\nu) \delta_{g,b},
\end{align}
 Let us define the following relation,
\begin{align}
    p(g|\mu,\beta,\nu)=\sum_{a,\alpha}p^*(a|\alpha,\mu)p(\alpha|\beta,\nu)p(g|a,\beta,\nu).
\end{align}
With this definition, the sum on the L.H.S. of the Inequality (\ref{Eq:12}) simplifies to $\sum_{\mu gb\beta \nu}w(b,\beta)p(\nu) \text{Tr} [\mathsf{C}_{b|\beta}Q_\mu^*]  p(g|\mu,\beta,\nu) \delta_{g,b}$. Notice that this expression is similar to $p(\{\mathcal{X}_{\beta}\})$. However, it is not in the most general form as $Q_\mu^*$ does not depend on $\nu$. Hence, the sum is never larger than the optimal success probability in QCD  game scenario played with access to single testers, $p(\{\mathcal{X}_{\beta}\})$. Therefore, we have 
\begin{align}\label{Eq:13}
    [1+\mathcal{R}(\{\mathsf{T}_{\alpha}\})]p(\{\mathcal{X}_{\beta}\})\geq \sum_{\mu,g,a,b,\alpha,\beta,\nu}w(b,\beta)p(\nu) p(\alpha|\beta,\nu) \text{Tr} [\mathsf{C}_{b|\beta}T_{a|\alpha}]  p(g|a,\beta,\nu) \delta_{g,b}.
\end{align}
The above inequality is valid for all the probabilities denoted as $p(\nu), p(\alpha|\beta,\nu),p(g|a,\beta,\nu)$. Therefore, the inequality will remain valid when we take maximization over all such probabilities. With this maximization, R.H.S. of Inequality (\ref{Eq:13}) is equal to the optimal success probability in QCD game with the incompatible testers,  $p(\{\mathcal{X}_{\beta}\},\{ \mathsf{T}_{\alpha}\})$. Eventually, we are left with the following bound,
\begin{equation}
	1+\mathcal{R}(\{\mathsf{T}_{\alpha}\}) \geq \max_{\{\mathcal{X}_{\beta}\}}\frac{  p(\{\mathcal{X}_{\beta}\},\{ \mathsf{T}_{\alpha}\})}{  p(\{\mathcal{X}_{\beta}\})}.
\end{equation}
Hence, we find that $1+\mathcal{R}(\{\mathsf{T}_{\alpha}\})$ is always greater than or equal to the maximal relative performance between the two strategies, while maximised over all possible ensembles. 
\subsection{Dual SDP}
\noindent The Lagrangian associated with our primal SDP {(defined in Eqs. \eqref{Eq:A26}- \eqref{Eq:A29})} is the following,
\begin{align}
    \mathcal{L}=&s+\sum_{a,\alpha}{\rm Tr}\Big[\omega_{a\alpha}\left(T_{a|\alpha}-\sum_{\Vec{a}} d_{\Vec{a}}(a|\alpha) \Tilde{Q}_{\Vec{a}}\right)\Big]-{\rm Tr}\Big[X\left(\mathds{1}\otimes \tilde{\Theta}^{(n)}- \sum_{\Vec{a}}\Tilde{Q}_{\Vec{a}} \right)\Big]\nonumber
    \\
    & -{\rm Tr}\sum_{\Vec{a}}x_{\Vec{a}}\Tilde{Q}_{\Vec{a}}
    -\text{Tr}[y^{(k)}(\mathds{1}_{(2k-3)} \otimes \tilde{\Theta}^{(k-1)} - \text{Tr}_{(2k-2)}[\tilde{\Theta}^{(k)}])]\Big).
\end{align}
where we have introduced the associated dual variables. By regrouping, we can write
\begin{align}
    \mathcal{L}=&s(1- \text{Tr}[X(\mathds{1}\otimes \Theta^{(n)})] - \sum_{k=2}^{n}\text{Tr}[y^{(k)}(\mathds{1}_{(2k-3)} \otimes \Theta^{(k-1)}- \mathds{1}_{(2k-2)} \otimes \Theta^{(k)})] ) \\ &+{\rm Tr}\sum_{a,\alpha}\omega_{a\alpha}T_{a|\alpha}+{\rm Tr}\sum_{\Vec{a}}\Big[X- \sum_{a,\alpha}\omega_{a\alpha}d_{\Vec{a}}(a|\alpha)-x_{\Vec{a}}\Big] \Tilde{Q}_{\Vec{a}}.
\end{align}
The Lagrangian is made independent of the primal variables when $\text{Tr}[X(\mathds{1}\otimes \Theta^{(n)})] + \sum_{k=2}^{n}\text{Tr}[y^{(k)}(\mathds{1}_{(2k-3)} \otimes \Theta^{(k-1)}- \mathds{1}_{(2k-2)} \otimes \Theta^{(k)})]  = 1$, and $X=\sum_{a,\alpha}\omega_{a\alpha}d_{\Vec{a}}(a|\alpha)+ x_{\Vec{a}}$ $\forall \Vec{a}$.
This enables us to write down the dual SDP as,
\begin{align}
    1+\mathcal{R}(\{\mathsf{T}_{\alpha}\}) =& \max_{y,X,\{\omega_{a\alpha}\}}  {\rm Tr}\left[\sum_{a,\alpha}\omega_{a\alpha}T_{a|\alpha}\right] \\
	& \text{s.t.}\:\:X\geq \sum_{a,\alpha}\omega_{a\alpha}d_{\Vec{a}}(a|\alpha),\:\: \omega_{a\alpha}\geq 0\\
     & \text{Tr}[X(\mathds{1}\otimes \Theta^{(n)})]  + \sum_{k=2}^{n}\text{Tr}[y^{(k)}(\mathds{1}_{(2k-3)} \otimes \Theta^{(k-1)}- \mathds{1}_{(2k-2)} \otimes \Theta^{(k)})]=1.\label{condine}
\end{align}
We can find solutions for $y^{(k)}$s for the relation in Eq. \eqref{condine} which result in satisfying $\text{Tr}[X(\mathds{1}\otimes \Theta^{(n)})] \leq 1$. 
Therefore, the final dual SDP is of the form,
\begin{align}
    1+\mathcal{R}(\{\mathsf{T}_{\alpha}\}) =& \max_{X,\{\omega_{a\alpha}\}}  {\rm Tr}\left[\sum_{a,\alpha}\omega_{a\alpha}T_{a|\alpha}\right] \label{Eq:A42}\\
	& \text{s.t.}\:\:X\geq \sum_{a,\alpha}\omega_{a\alpha}d_{\Vec{a}}(a|\alpha) \nonumber\\
     & \omega_{a\alpha}\geq 0,\:\: \text{Tr}[X(\mathds{1}\otimes \Theta^{(n)})]\leq 1 \nonumber.
\end{align}
{Equivalence between the primal and dual happens if strong duality exists, i.e., a strictly feasible solution exists for a dual problem. One such solution is $X=\mathds{1}/D$, and 
$\omega_{a\alpha}=\gamma \mathds{1}$ for $\gamma$ such that $1/(mD)>\gamma>0$ for $m$ $\in \mathds{R}^+$, where $D = \prod_{j=0}^{n}d_{2n-(2j+1)}$.} 
\subsection{Upper bound}
Consider that the optimal dual variables $X^*$, and $\omega_{a\alpha}^*$, then we have from Eq. \eqref{Eq:A42},
\begin{align}\label{Eq:dual-sdp}
    1+\mathcal{R}(\{\mathsf{T}_{\alpha}\}) = & {\rm Tr}\left[\sum_{a,\alpha}\omega_{a\alpha}^*T_{a|\alpha}\right] \\
	& \text{s.t.}\:\: X^*\geq \sum_{a,\alpha}\omega_{a\alpha}^*d_{\Vec{a}}(a|\alpha) \nonumber\\
     &  \omega^*_{a\alpha}\geq 0,\:\: \text{Tr}[X^*(\mathds{1}\otimes \Theta^{(n)})]\leq 1\nonumber.
\end{align}
Let us introduce three auxiliary variables to have a connection with QCD game, 
\begin{align}\label{dual-variable}
    \mathsf{C}_{a|\alpha}^*=\frac{\omega_{a\alpha}^*}{\ell^* q^*(a,\alpha)},\quad
   {\rm s.t.}\quad {\rm Tr}[\mathsf{C}_{a|\alpha}^*]=\Bar{D},
   \:\:{\rm and}\:\: q^*(a,\alpha)={\rm Tr}[\omega_{a\alpha}^*]\Big/\sum_{a,\alpha}{\rm Tr}[\omega_{a\alpha}^*],
\end{align}
{where $\Bar{D} = \prod_{j=0}^{n}d_{2n-(2j+1)}$ with $d_x$ is the dimension of the system labeled ``$x$" and $\{q^*(a,\alpha)\}$ is normalized probability distribution. This implies $\ell^*=(1/\Bar{D})\sum_{a,\alpha}{\rm Tr}[\omega_{a\alpha}^*]$}.  Then, Eq.(\ref{Eq:dual-sdp}), can be rewritten as 
\begin{align}
    1+\mathcal{R}(\{\mathsf{T}_{\alpha}\}) = \ell^* \sum_{a,\alpha}q^*(a,\alpha){\rm Tr}[\mathsf{C}_{a|\alpha}^*T_{a|\alpha}]
\end{align}
Let us assume that Alice will play the QCD game with following strategy. After receiving the ensemble of combs $\{\mathcal{X}^*_{\beta}\}_\beta$ with $\mathcal{X}_{\beta}^* = \{ \mathsf{C}_{b|\beta}^*,w^*(b|\beta)\}_{b}$, she plays the game with $p(\nu)=\delta_{\nu,0}$ (i.e., sets $\nu=0$), $p(\alpha|\beta,\nu=0)=\delta_{\alpha,\beta}$ (i.e., applies $\mathsf{T}_\beta$ given $\beta$) and $p(g|a,\beta,\nu=0)=\delta_{g,b}$ (i.e., guesses $b=a=g$ after getting outcome $a$). 
Choosing this sub-optimal strategies, one gets,
\begin{align}
    p(\{\mathcal{X}^*_{\beta}\},\{ \mathsf{T}_{\alpha}\})&\geq\sum q^*(b,\beta)\delta_{\nu,0}\delta_{\alpha,\beta}{\rm Tr}[\mathsf{C}_{b|\beta}^*T_{a|\alpha}]\delta_{g,b}\delta_{a,g},\nonumber\\
    &=\sum_{a,\alpha}q^*(a,\alpha){\rm Tr}[\mathsf{C}_{a|\alpha}^*T_{a|\alpha}]\nonumber\\
    &=(1/\ell^*)[1+\mathcal{R}(\{\mathsf{T}_{\alpha}\})].
\end{align}
Now, from the constraint of above dual SDP, we find following relation for the current game that 
\begin{align}
    X^*\geq \sum_{b,\beta}d_{\Vec{b}}(b|\beta) \ell^*q^*(b,\beta)\mathsf{C}_{b|\beta}^*.
\end{align}
Then, multiplying both sides by an arbitrary Tester $\Tilde{Q}_{\Vec{b}}$, summing over $\Vec{b}$ and taking trace, we get
\begin{align}
  {\rm Tr}  \left[X^*\sum_{\Vec{b}}\Tilde{Q}_{\Vec{b}}\right]\geq \sum_{b,\beta,\Vec{b}}d_{\Vec{b}}(g|\beta)\delta_{b,g} \ell^*q^*(b,\beta){\rm Tr}\left[\mathsf{C}_{b|\beta}^*\Tilde{Q}_{\Vec{b}}\right].
\end{align}
Notice that due to normalization of an $n$-slot tester, the lhs is exactly equal to unity. Also, the above equation should hold if we maximize over all $\Tilde{Q}_{\Vec{b}}$. This implies that 
\begin{align}\label{Eq:14}
  \frac{1}{\ell^*}\geq \max_{\Tilde{Q}_{\Vec{b}}}\sum_{b,\beta,\Vec{b}} q^*(b,\beta){\rm Tr}\left[\mathsf{C}_{b|\beta}^*\Tilde{Q}_{\Vec{b}}\right]d_{\Vec{b}}(g|\beta)\delta_{b,g}.
\end{align}
One can see the resemblance of rhs of above equation with the success probability in QCD game with only single measurement. In particular, we prove below in (\ref{proof-ineq}) that  $(1/\ell^*)\geq p(\{\mathcal{X}_{\beta}^*\})$. Therefore, it implies
\begin{equation}
	  1+\mathcal{R}(\{\mathsf{T}_{\alpha}\}) \leq \frac{  p(\{\mathcal{X}_{\beta}^*\},\{ \mathsf{T}_{\alpha}\})}{  p(\{\mathcal{X}_{\beta}^*\})}.
\end{equation}
Hence, we arrive at the concluding from both upper and lower bound of $1+\mathcal{R}(\{\mathsf{T}_{\alpha}\})$ that
\begin{equation}
	\max_{\{\mathcal{X}_{\beta}\}} \frac{  p(\{\mathcal{X}_{\beta}\},\{ \mathsf{T}_{\alpha}\})}{  p(\{\mathcal{X}_{\beta}\})}=1+\mathcal{R}(\{\mathsf{T}_{\alpha}\}).
\end{equation}
\subsubsection{Proof of $(1/\ell^*)\geq p(\{\mathcal{X}_{\beta}^*\})$}\label{proof-ineq}
Let us recall the success probability defined below,
\begin{align}
    p(\{\mathcal{X}_{\beta}\}) = \max_{\mathcal{S}_c} \sum_{b,\beta,a,g,\nu}  w(b,\beta)p(\nu)  \text{Tr}[\mathsf{C}_{b|\beta}Q_{a|\nu}]p(g|a,\beta,\nu)\delta_{g,b}.
\end{align}
We already know that it is possible to write $p(g|a,\beta,\nu)=\sum_{\Vec{b}}p(\Vec{b}|a,\nu)d_{\Vec{b}}(g|\beta)$, then we can rewrite the below expression as 
\begin{align}
    \sum_{b,\beta,a,g,\nu}  w(b,\beta)p(\nu)  \text{Tr}[\mathsf{C}_{b|\beta}Q_{a|\nu}]p(g|a,\beta,\nu)\delta_{g,b}=\sum_{b,\beta,\Vec{b}}w(b,\beta)  \text{Tr}[\mathsf{C}_{b|\beta}\Tilde{Q}_{\Vec{b}}]d_{\Vec{b}}(g|\beta)\delta_{g,b},
\end{align}
where we introduce a new object defined as 
\begin{align}
    \Tilde{Q}_{\Vec{b}}:=\sum_{a,\nu}p(\nu)Q_{a|\nu}p(\Vec{b}|a,\nu).
\end{align}
Notice that for all values of $\Vec{b}$, $\Tilde{Q}_{\Vec{b}}\geq 0$ and also satisfies following 
\begin{align}
   \sum_{\Vec{b}} \Tilde{Q}_{\Vec{b}}&=\sum_{a,\nu,\Vec{b}}p(\nu)Q_{a|\nu}p(\Vec{b}|a,\nu)=\sum_{a,\nu}p(\nu)Q_{a|\nu}\sum_{\Vec{b}}p(\Vec{b}|a,\nu)\nonumber\\
   &=\sum_\nu p(\nu)\sum_a Q_{a|\nu}={  \sum_\nu p(\nu) \mathds{1}_{(2n-1)} \otimes \Xi^{(n)}_\nu } \nonumber = {  \mathds{1}_{(2n-1)} \otimes \Xi^{(n)},  }
\end{align}
{where $\Xi^{(n)}$ denotes a valid $n$-slot comb.}
It is evident now that this strategy with $\Tilde{Q}_{\Vec{b}}$ is as good as the strategy with $Q_{a|\nu}$. Therefore, we write equivalently, 
\begin{align}
  p(\{\mathcal{X}_{\beta}\}):=\max_{\Tilde{Q}_{\Vec{b}}}\sum_{b,\beta,\Vec{b}} w(b,\beta){\rm Tr}\left[\mathsf{C}_{b|\beta}\Tilde{Q}_{\Vec{b}}\right]d_{\Vec{b}}(g|\beta)\delta_{b,g}.
\end{align}
Using this in Eq. (\ref{Eq:14}), we arrive at the conclusion that 
\begin{align}
    \frac{1}{\ell^*}\geq p(\{\mathcal{X}_{\beta}^*\}).
\end{align}
\section{Alternative proof concept for convex weight related results}\label{conv-app}
It is easy to see that we can convert the Eq. \eqref{conic1} to an SDP. Using the definition of convex weight of tester assemblage $\{\mathsf{T}_\alpha\}_\alpha$ with $\mathsf{T}_\alpha\equiv \{T_{a|\alpha}\}_a$ in Eq. \eqref{conic0}, we get
\begin{align}
    T_{a|\alpha}-\mathcal{W}(\{\mathsf{T}_\alpha\})N_{a|\alpha}&=[1-\mathcal{W}(\{\mathsf{T}_\alpha\}]\sum_{\mu} p(a|\alpha,\mu) Q_{\mu},\nonumber\\
   & =\Bar{Q}_{a|\alpha},
\end{align}
where we have identified $[1-\mathcal{W}(\{\mathsf{T}_\alpha\})]\sum_{\mu} p(a|\alpha,\mu) Q_{\mu}:=\Bar{Q}_{a|\alpha}$. Now considering $s_w=1-\mathcal{W}(\{\mathsf{T}_\alpha\})$, we reach to an equivalent SDP formulation below for convex weight as
\begin{align}
    1-\mathcal{W}(\{\mathsf{T}_\alpha\})&=\max \quad s_w \\
    &{\rm s.t.}\quad \frac{1}{s_w}\left(T_{a|\alpha}-[1-s_w]N_{a|\alpha}\right)= \sum_{\mu} p(a|\alpha,\mu) Q_{\mu}.
\end{align}
Given this formulation, we can reach to the nicer form of SDP by similar logic from Appendix \ref{appen:1} and using the fact that $\{[1/(1-s_w)]\Bar{Q}_{a|\alpha}\}$ form a set of compatible testers also. Therefore, the prove of Eq. \eqref{conv-res} will follow similar to Appendix \ref{appen:1}.

\end{document}